\documentclass[12pt]{iopart}
\usepackage{graphicx,latexsym,stmaryrd,iopams}
 
\begin{document}

\title[QISM with anyonic grading]{Quantum Inverse Scattering Method with anyonic grading}

\author{M T Batchelor$^{1,2}$, A Foerster$^3$, X-W Guan$^1$, J Links$^4$ and H-Q Zhou$^5$}
\address{$^1$  
{\small Department of Theoretical Physics, Research School of Physical Sciences and Engineering, 
Australian National University, Canberra ACT 0200, Australia}}
\address{$^2$
{\small Mathematical Sciences Institute, Australian National University, Canberra ACT 0200, Australia}}
\address{$^3$
{\small Instituto de F\'{\i}sica da UFRGS,  Av.\ Bento Gon\c{c}alves, 9500,
                     Porto Alegre, 91501-970, Brasil}}
\address{$^4$
{\small Centre for Mathematical Physics, 
School of Physical Sciences, The University of Queensland, 4072, Australia\\}}
\address{$^5$
{\small Centre for Modern Physics, Chongqing University, Chongqing 400044, P.R. China}}

\eads{\mailto{Murray.Batchelor@anu.edu.au}, \mailto{angela@if.ufrgs.br}, 
\mailto{xwe105@rsphysse.anu.edu.au}, \mailto{jrl@maths.uq.edu.au}
and \mailto{hqzhou@cqu.edu.cn}}

\date{\today}

\begin{abstract}
We formulate the Quantum Inverse Scattering Method for the case of anyonic grading.
This provides a general framework for constructing integrable models describing 
interacting hard-core anyons. 
Through this method we reconstruct the known integrable model of hard
 core anyons associated with the $XXX$ model, and as a new application we construct the anyonic $t-J$ model. 
The energy spectrum
 for each model is derived by means of a generalisation of the 
 algebraic Bethe ansatz. The grading parameters
 implementing the anyonic signature give rise to sector-dependent
 phase factors in the Bethe ansatz equations.
\end{abstract}




\section{Introduction}
\label{sec1}

The development of the quantum inverse scattering method (QISM) \cite{QISM} 
led to the discovery of a number of quantum integrable models.
Applications of the QISM to physical systems such as the 
Bloch electron problem \cite{Bloch}, 
BCS models \cite{BCS} and Bose-Einstein condensates \cite{BEC} 
have opened up new applications of well developed mathematical techniques
to describe low-dimensional many-body physics \cite{Hubbardbook,Tbook,BGOS}.
However, the main objects of the QISM are restricted to spin and fermion models 
(or two-dimensional classical vertex models) which are closely related to representations of Lie
algebras and Lie superalgebras in finite Hilbert spaces. 
Although the Jordan-Wigner transformation 
can be implemented between different statistics languages, such as
the transmutation from bosons to fermions \cite{JW2,JW3}, 
little attention has been paid to integrable models with generalized statistics. 
Some integrable lattice models with anyon-like commutation relations
have been constructed \cite{LIM,c-HM}. The anyonic statistical
parameters result in global phase factors acting as a gauge potential
in the Bethe ansatz equations. These phase factors lead to magnetic
flux-like effects. However, if anyonic commutation relations are
imposed on the $1$D continuum quantum gases, the dynamical interaction
and anyonic statistical interaction are inextricably intertwined
\cite{Anyon-1}, resulting in quite different low
energy properties and statistical effects
\cite{Anyon-2}
than the standard $1$D Bose and Fermi gases \cite{Tbook}.

It is now well understood that the integrable quantum fermion models
can be treated by the graded QISM where the Grassmann parity is 
adapted to fit the anticommuting property of fermions \cite{Kulish,Wad1,GM}.  It is
natural to ask whether one can modify the usual QISM to treat other
models with different statistics, like fractional statistics and
anyonic statistics \cite{Haldane,Wu}.
Here we show that integrable lattice models of hard-core anyons can be
systematically constructed via the Yang-Baxter equation with $U(1)$
Abelian group-like grading called anyonic grading. This is a
generalization of $\mathbb{Z}_2$ grading to a continuous $U(1)$ grading function. The
anyonic grading has a similar signature as the colour grading invented
by Rittenberg and Wyler \cite{RW} but is not fully equivalent.

In this paper, we generalize the QISM
to the anyonic grading QISM which can be used to construct quantum
integrable models describing hard-core anyons. As a first example of
the anyonic QISM, we reconstruct the integrable $XXX$ type model and
its exact solution which has previously been studied in \cite{LIM} via
the co-ordinate Bethe ansatz. As a new application, we then  consider
the anyonic su(3) $t-J$ model
with the Hamiltonian written in terms of hard-core anyon operators,
and exactly solve it by the algebraic Bethe ansatz. This gives the
energy spectrum in terms of the Bethe ansatz equations. The anyonic 
grading functions appear in the Bethe ansatz equations resulting in
anyonic signature.  

Our motivation for developing the QISM with anyonic grading is that it
leads to wider application for studying exactly solvable
one-dimensional lattice models than the co-ordinate Bethe ansatz
approach, which has previously been discussed in
\cite{LIM,c-HM}. There are two main reasons for this. 
The first is that in the co-ordinate Bethe ansatz approach the anyonic
statistics are introduced through the use of canonical operators which
are anyonic deformations of the familiar canonical fermion operators,
that can in principle be constructed through Jordan-Wigner type
transformations. The algebraic approach we describe is more general and
does not depend on the existence of such a representation of the Hilbert
space of states. Secondly, the algebraic approach is more accessible for
extending the analysis toward the computation of form factors and
correlation functions where Jordan-Wigner transformations cannot be
directly implemented due to the non-locality of the transformations. In
the QISM approach, the anyonic grading parameters will directly arise in
the correlation functions through the commutation relations between
creation  
fields and annihilation fields in the algebraic Bethe ansatz scheme.
Moreover, the norms of the Bethe ansatz wave functions
can be represented as determinants \cite{QISM,form1,form2}. 
Other interesting
applications of anyonic grading to quantum inverse problems
\cite{QIP1,QIP2} and $R$-operator representation \cite{QIP1,Umeno} 
should be straightforward.  For example, for the
anyonic $t-J$ model we will construct below, the form factors and
correlation functions can be determined following the procedure used
for the Heisenberg XXX model \cite{form1,form2} and for the supersymmetric $t-J$ model \cite{zyz}.

This paper is organized as follows. In section \ref{sec2} we introduce
some basic concepts for the generalized   grading and present the anyonic
grading QISM. We give an explicit expression for the Hamiltonian and
derive the Bethe ansatz solution for the $XXX$ model of hard core
anyons in section \ref{sec3}. In section \ref{sec4} the $t-J$ model of
hard core anyons is constructed and the exact solution is obtained by
the algebraic Bethe ansatz.  Concluding remarks are given in section
\ref{sec5}.

\section{QISM with anyonic grading}
\label{sec2}

The standard colour algebras \cite{RW,C-S} are defined through the notion of colour
graded vector spaces. The colour
structures are a generalization of supersymmetric structures in that
the grading with respect to ${\mathbb Z}_2$ is generalized to an arbitrary
Abelian group $\Gamma$. For anyonic grading with the Abelian group being $U(1)$, we
can directly define operations in a parallel way to colour
grading. Due to the grading being associated with $U(1)$, we always consider cases where the underlying fields for the vector spaces are ${\mathbb C}$.   

Letting $U,\,V$ denote complex vector spaces with bases $\{u^i\},\,\{v^j\}$, 
the anyonic
permutation operator  
$P:~U\otimes V~\rightarrow ~V\otimes  U$ is
defined by the action on the basis vectors
\begin{equation}
P(u^i\otimes v^j)=w(i,j)(v^j\otimes u^i)
\label{P}
\end{equation}
where $w(i,j)\in\,U(1)$ are the anyonic grading parameters. The inverse operator $P^{-1}:~V\otimes U~\rightarrow ~U\otimes  V$
has the action
\begin{eqnarray*}
P^{-1}(v^j \otimes u^i)&=& \tilde{w}(j,i)(u^i\otimes v^j) \\
&=& w^{-1}(i,j)(u^i\otimes v^j). 
\end{eqnarray*}
This implies that in the special case where $U=V$, which occurs when dealing with indistinguishable particles,  the anyonic grading parameters must possess the symmetry 
$$w(i,j)=w(j,i). $$ 
Other than this there are no constraints imposed on the choice of the $w(i,j)$.
A significant difference between our formulation of anyonic grading and that of colour grading is that for colour grading the constraint 
$$w(i,i)=\pm 1 $$ 
is imposed, whereas for anyonic grading we relax this condition. 
For each choice of anyonic grading it is natural to also define the dual grading with permutation operator $P^*$ acting as 
$$P^*(u^i\otimes v^j)=w^{-1}(i,j)(v^j\otimes u^i). $$

Using the
fundamental basis of linear operators $\{e^i_j\}$ acting on $U$ and $\{f^k_l\}$ acting on $V$
such that 
$$ e^i_j u^m=\delta^m_j u^i,\qquad\quad f^k_l v^n= \delta^n_l v^k,$$ 
 the basis for  the anyonic  graded tensor product ${\rm End} (U)\otimes_{\rm a} {\rm End} (V)$ is defined by 
\begin{equation}
e^i_{j}\otimes _{\rm a}f^k_{l} =w(j,k)w^{-1}(j,l)e^i_{j}\otimes f^k_{l}.
\label{def}
\end{equation}
We say that the basis operator $e^i_j\otimes_{\rm a} f^k_l$ is {\it even} if 
$$w(j,k)w^{-1}(j,l)=1 $$
and more generally an operator is even if it is a linear combination of even basis operators. 
If we introduce a third vector space $W$ with basis $\{g^r_s\}$ then it follows from (\ref{def}) that the anyonic graded tensor product is associative:
$$ (e^i_{j}\otimes _{\rm a}f^k_{l})\otimes_{\rm a} g^r_s 
= e^i_{j}\otimes _{\rm a}(f^k_{l}\otimes_{\rm a} g^r_s ). $$

The basis for the opposite anyonic graded tensor product ${\rm End} (V)\otimes_{\rm a} {\rm End} (U)$  is defined in terms of $P^{-1}$
\begin{eqnarray*}
f^k_l\otimes _{\rm a}e^i_j&=&\tilde{w}(l,i)\tilde{w}^{-1}(l,j)f^k_l\otimes e^i_j \\     
&=&w^{-1}(i,l)w(j,l) f^k_l\otimes e^i_j .
\end{eqnarray*}
Now we define the twist map $T: {\rm End} (V)\otimes_{\rm a} {\rm End} (U) \rightarrow 
{\rm End} (U)\otimes_{\rm a} {\rm End} (V) $. 
It is defined through the inverse anyonic permutation operator and its dual as 
\begin{eqnarray*}
T(f^k_l \otimes_{\rm a} e^i_j)&=& (P^*)^{-1}(f^k_l \otimes_{\rm a} e^i_j)P^{-1} \\
&=&w(j,l)w^{-1}(i,l)(P^*)^{-1}(f^k_l \otimes e^i_j)P^{-1} \\
&=&w(j,l)w^{-1}(i,l)w(i,k)w^{-1}(j,l)(e^i_j \otimes f^k_l) \\
&=&w(i,k)w(j,l)w^{-1}(i,l)w^{-1}(j,k)(e^i_j \otimes_{\rm a} f^k_l). 
\end{eqnarray*}
Through the twist map $T$ and the usual matrix multiplication $m_U:{\rm End} (U)\otimes {\rm End} (U) \rightarrow {\rm End} (U)$ the anyonic graded tensor product multiplication is formally defined as 
\begin{eqnarray*}
&&(e^i_j\otimes_{\rm a} f^k_l)(e^p_q\otimes_{\rm a}f^r_s) \\
&&\qquad= (m_U \otimes m_V) ({\rm id} \otimes T \otimes {\rm id})
(e^i_j\otimes_{\rm a} f^k_l\otimes_{\rm a}e^p_q\otimes_{\rm a}f^r_s) \\
&&\qquad = w(p,k)w(q,l)w^{-1}(p,l)w^{-1}(q,k)(m_U \otimes m_V)(e^i_j\otimes_{\rm a} e^p_q \otimes_{\rm a}   f^k_l \otimes_{\rm a}f^r_s)  \\
&&\qquad = w(p,k)w(q,l)w^{-1}(p,l)w^{-1}(q,k)(e^i_je^p_q \otimes_{\rm a} f^k_l  f^r_s) \\
&&\qquad = w(p,k)w(q,l)w^{-1}(p,l)w^{-1}(q,k)(e^i_q \otimes_{\rm a} f^k_s).     
\end{eqnarray*}
On the other hand working directly with the definition (\ref{def}) we have
\begin{eqnarray}
(e^i_j\otimes_{\rm a} f^k_l)(e^p_q\otimes_{\rm a}f^r_s)
&=&w(j,k)w^{-1}(j,l)w(q,r)w^{-1}(q,s)(e^i_j\otimes f^k_l)(e^p_q\otimes f^r_s)\nonumber  \\
&=&w(p,k)w(q,l)w^{-1}(p,l)w^{-1}(q,s) e^i_q\otimes f^k_s \nonumber \\
&=&w(p,k)w(q,l)w^{-1}(p,l)w^{-1}(q,k) e^i_q \otimes_{\rm a} f^k_s
\label{product}
\end{eqnarray}
which shows that the definitions for the anyonic graded tensor product and its multiplication are consistent.

The ${\mathbb Z}_2$ graded QISM was set up in \cite{GM}. Here we establish an
analogous anyonic  graded QISM.  A matrix $R(\lambda )$ is said
to fulfill the Yang--Baxter equation (YBE) with anyonic  grading if the identity
\begin{eqnarray}
& &\left(I\otimes _{{\rm a}}\stackrel{\vee}{R}(\lambda -\mu)\right)\left(\stackrel{\vee}{R}(\lambda )\otimes _{{\rm a}}I\right)\left(I\otimes _{{\rm a}}\stackrel{\vee}{R}(\mu)\right)\nonumber\\
& &=
\left(\stackrel{\vee}{R}(\mu )\otimes _{{\rm a}}I\right)\left(I\otimes _{{\rm a}}\stackrel{\vee}{R}(\lambda )\right)\left(\stackrel{\vee}{R}(\lambda -\mu )\otimes _{{\rm a}}I\right)\label{CYBE}
\end{eqnarray}
acting on $V_1\otimes _{{\rm a}}V_2\otimes _{{\rm a}}V_3$ holds.
We will impose that the $\stackrel{\vee}{R}$-matrix is chosen to be even. 
Thus the YBE with anyonic  grading can be written in component form
\begin{eqnarray}
\stackrel{\vee}{R}(\lambda-\mu)^{{\scriptstyle a_2a_3}}_{{\scriptstyle c_2c_3}}
\stackrel{\vee}{R}(\lambda)^{{\scriptstyle a_1c_2}}_{{\scriptstyle b_1d_2}}
\stackrel{\vee}{R}(\mu)^{{\scriptstyle d_2c_3}}_{{\scriptstyle b_2b_3}}=
\stackrel{\vee}{R}(\mu)^{{\scriptstyle a_1a_2}}_{{\scriptstyle c_1c_2}}
\stackrel{\vee}{R}(\lambda)^{{\scriptstyle c_2a_3}}_{{\scriptstyle d_2b_3}}
\stackrel{\vee}{R}(\lambda-\mu)^{{\scriptstyle c_1d_2}}_{{\scriptstyle b_1b_2}}.\label{CYBEi}
\end{eqnarray}
The summation convention is implied for the repeated indices
$a_j,b_j,c_j$ and $d_j$. We notice that despite the fact that the
tensor product (\ref{CYBE}) carries the anyonic grading, there are no
extra anyonic grading parameters in (\ref{CYBEi}) compared to the
standard one.
This is because we consider the case where the $R$-matrix  is even. 

With the help of the anyonic permutation operator
(\ref{P}) we may, from (\ref{CYBE}), prove the anyonic  graded YBE in the form
\begin{equation}
R_{12}(\lambda -\mu)R_{13}(\lambda)R_{23}(\mu)=R_{23}(\mu)R_{13}(\lambda )R_{12}(\lambda-\mu), \label{CYBE2}
\end{equation}
where $R(\lambda )=P\stackrel{\vee}{R}(\lambda)$. Similarly, in component form it reads
\begin{eqnarray}
& &R(\lambda -\mu)^{{\scriptstyle f_1e_2}}_{{\scriptstyle c_1c_2}}
{R}(\lambda)^{{\scriptstyle c_1f_3}}_{{\scriptstyle b_1c_3}}{R}(\mu)^{{\scriptstyle c_2c_3}}_{{\scriptstyle b_2b_3}}
w(b_1,c_2)w^{-1}(c_1,c_2)\nonumber\\
& &
{R}(\mu)^{{\scriptstyle f_1f_3}}_{{\scriptstyle c_1c_3}}
{R}(\lambda )^{{\scriptstyle e_2c_3}}_{{\scriptstyle c_2b_3}}
R(\lambda-\mu)^{{\scriptstyle c_1c_2}}_{{\scriptstyle b_1b_2}}
w(c_1,e_2)w^{-1}(f_1,e_2).
\end{eqnarray}
If we choose the spaces $V_1,V_2$ as auxiliary spaces, the space $V_3$
as the quantum space, then letting $L_n(\lambda)=R_{0n}(\lambda)$ the
anyonic graded YBE (\ref{CYBE2}) becomes
\begin{equation}
R_{00^{'}}(\lambda -\mu)R_{0n}(\lambda)R_{0^{'}n}(\mu)=R_{0^{'}n}(\mu)R_{0n}(\lambda )R_{00^{'}}(\lambda-\mu), \label{CYBE3}
\end{equation}
or equivalently
\begin{equation}
\stackrel{\vee}{R}(\lambda -\mu)L_n(\lambda)\otimes _{{\rm a}}L_n(\mu)=
L_n(\mu)\otimes _{{\rm a}}L_n(\lambda)\stackrel{\vee}{R}(\lambda-\mu).\label{CYBE4}
\end{equation}
In component form
\begin{eqnarray}
&&\stackrel{\vee}{R}(\lambda-\mu)^{{\scriptstyle a_1a_2}}_{{\scriptstyle c_1c_2}}
L_n(\lambda)^{{\scriptstyle a_1a_n}}_{{\scriptstyle b_1r_n}}
L_n(\mu)^{{\scriptstyle c_2r_n}}_{{\scriptstyle b_2b_n}}w(b_1,c_2)w^{-1}(c_1,c_2)=\nonumber\\
& &
L_n(\mu)^{{\scriptstyle a_1a_n}}_{{\scriptstyle c_1r_n}}
L_n(\lambda)^{{\scriptstyle a_2r_n}}_{{\scriptstyle c_2b_n}}
\stackrel{\vee}{R}(\lambda-\mu)^{{\scriptstyle c_1c_2}}_{{\scriptstyle
    b_1b_2}}w(c_1,a_2)w^{-1}(a_1,a_2). \label{AYBA}
\end{eqnarray}
The Yang-Baxter algebra (\ref{AYBA}) with anyonic grading naturally
provides a set of anyonic commutation relations for  interacting hard-core
anyons \footnote{A similar form of the Yang-Baxter algebra was given
in the context of braided quantum YBE \cite{Kundu2}.}.
As a consequence, the anyonic grading gives rise to sector-dependent phase
factors in the Bethe ansatz solution. The subtlety of the anyonic
grading parameters is seen clearly in the anyonic su(3) $t-J$ 
model discussed in section \ref{sec4}.

Let us define the monodromy matrix $T(\lambda)$ as the matrix product over the Lax operators on all sites of the lattice, i.e.
\begin{equation}
T(\lambda )=L_N(\lambda)L_{N-1}(\lambda)\cdots L_1(\lambda).
\end{equation}
Here 
$T(\lambda)$ is a quantum operator valued matrix that acts
non-trivially in the anyonic tensor product of a whole quantum space
of the lattice and satisfies the global anyonic  graded YBE 
\begin{equation}
R(\lambda -\mu)T(\lambda)\otimes_{{\rm a}}T(\mu)=T(\mu)\otimes_{{\rm a}}T(\lambda)R(\lambda -\mu). \label{gYBA}
\end{equation}
Consequently the transfer matrix $\displaystyle \tau(\lambda)={\rm atr}[T(\lambda)]=
\sum ^{n}_{a=1}w(a,a)^{-1}T(\lambda)^a_a$ forms a commuting family for all values of the spectral parameters. 
Here ${\rm atr}$ is the anyonic  graded
trace carried out in the auxiliary space with $n$ the dimension of the
auxiliary space. It follows that the transfer matrix can be considered
as the generating functional of the Hamiltonian and of an infinite
number of higher conservation laws of the model.

\section{The XXX model of hard core anyons}
\label{sec3}

As a first step we 
consider the integrable hard-core anyon model with the Hamiltonian 
\begin{eqnarray}
H&=&\eta^{-1}\sum^{L}_{j=1}\left(a^{\dagger}_{j+1}a_j+a^{\dagger}_ja_{j+1}+2n_{j+1}n_j-2n_{j}\right)\nonumber\\
&=&\eta^{-1}\sum^{L}_{j=1}\left(qa_ja^{\dagger}_{j+1}+q^{-1}a_{j+1}a^{\dagger}_j+2n_{j+1}n_j-2n_{j}\right)
 \label{model1}
\end{eqnarray}
where 
the operator $n_j=a_j^{\dagger}a_j$ is the number operator of
hard-core anyons and $a^{\dagger}_j$ and $a_j$ are the creation
and annihilation hard-core anyon operators satisfying the 
commutation relations
\begin{eqnarray}
& &\left\{a_j,a_j\right\}=\left\{a_j^{\dagger},a_j^{\dagger}\right\}=0~~
\left\{a_j,a^{\dagger}_j\right\}=1\\
& &
a^{\dagger}_ia_j=qa_ja^{\dagger}_i,~~a^{\dagger}_ja_i=q^{-1}a_ia^{\dagger}_j.\label{Lax}\\
& &a^{\dagger}_ja^{\dagger}_i=qa^{\dagger}_ia^{\dagger}_j,~~
a_ja_i=qa_ia_j.\label{comm-m1}
\end{eqnarray}
Here we assume $i~>j$ with $\left\{~\right\}$ denoting the anticommutator
as usual.  The Hamiltonian (\ref{model1}) reveals an anyonic signature
when particles interchange. The above on-site commutation relations
are indicative of hard-core particle behaviour.  We note that the hard-core anyons
\cite{JW2} preserve the Pauli exclusion principle whereas the off-site
ones have a free anyonic parameter when two particles exchange their
positions. This model (more specifically the $XXZ$ generalisation) was previously 
solved in \cite{LIM} using the co-ordinate Bethe ansatz. Below, we will confirm that 
this model arises through the anyonic QISM with the same solution as obtained 
via the algebraic Bethe ansatz. 

Consider the quantum $R$-matrix of the XXX model
\begin{equation}
\stackrel{\vee}{R}(\lambda)=\left(\begin{array}{cccc}
\lambda+\eta &0&0&0\\
0&\eta &\lambda &0\\
0&\lambda & \eta &0\\
0&0&0&\lambda+\eta
\end{array}
\right),
\end{equation}
where $\eta$ is a quasiclassical parameter. If we choose the anyonic  parity 
\begin{equation}
w(1,1)=w(1,2)=w(2,1)=1;~~w(2,2)=q,
\end{equation}
the anyonic grading  Lax operator on site $j$ is given by 
\begin{equation}
L_j(\lambda)=\left(\begin{array}{cc}
\lambda +\eta(1-n_j)&\eta a^{\dagger}_j\\
\eta a_j &\lambda +(\lambda (q-1)+q\eta )n_j
\end{array}
\right).\label{LM}
\end{equation}
 It is easy to check that the Lax operator (\ref{LM}) does satisfy the
 anyonic grading YBE (\ref{CYBE4}) with the commutation relations (\ref{comm-m1}). 
 As a consequence, the monodromy
 matrix generates the global anyonic grading YBE (\ref{gYBA}). Then
 the integrals of motion of the model can be obtained
 from the expansion of the transfer matrix in the spectral parameter
 $\lambda$. Explicitly,
\begin{equation}
\tau (\lambda)=(1+H\lambda+\cdots )\tau(0),
\end{equation}
where the Hamiltonian reads
\begin{equation}
H=\sum ^{N-1}_{i=1}H_{ii+1}+H_{N1}.\label{Ham}
\end{equation}
Here 
\begin{eqnarray}
H_{jj+1}& = &L_{j+1}(0)L^{'}_j(0)L^{-1}_j(0)L_{j+1}^{-1}(0),\nonumber\\
H_{N1}& = &{\rm atr}\left(L^{'}_N(0)L^{-1}_N(0)L_{1}(0)\right)\nonumber\\
 & =&\eta ^{-1} {\rm atr}\left(L^{'}_N(0)L_{1}(0)\right)P^{-1}_{1N}.\label{Hamd}
\end{eqnarray}
The properties
\begin{equation}
P^{-1}_{12}\stackrel{1}{X}P_{12}=\stackrel{2}{X},~~P^{-1}_{0j}P_{0k}P_{0j}=P_{jk}
\end{equation}
are applied in the above derivation. These properties imply a constraint on the grading function such that $w(\alpha, \beta)=w(\beta, \alpha)$.
After a lengthy algebraic calculation, the explicit expression for the Hamiltonian density and the boundary 
terms is given by (up to a constant)
\begin{eqnarray}
H_{jj+1}& = &(1-n_{j+1})(1-n_j)+n_{j+1}n_j+a^{\dagger}_{j+1}a_j+a^{\dagger}_ja_{j+1},\\
H_{N1}& = &(1-n_{1})(1-n_{N})+n_{1}n_{N}+a^{\dagger}_{1}a_N+a^{\dagger}_Na_{1},
\end{eqnarray}
which preserve  the periodic boundary condition for the model (\ref{model1}). 
To keep the Hamiltonian (\ref{model1}) hermitian, we restrict ourselves to 
$q^{*}=q^{-1}$, where the superscript $*$ denotes complex conjugation.
We 
remark that this model covers the hard-core boson model and the fermion model as
special choices of the anyonic  gradings. For example, for $q =1$ the model
corresponds to a hard-core boson XXX model. Using the Matsubara and
Matsuda transformations \cite{MM,JW2}, this hard-core model becomes the
standard XXX vertex model. For $q=-1$ it is the $su(2)$ $XXX$
fermion chain.  

After performing the standard algebraic Bethe ansatz, the transfer matrix
eigenvalues are of the form
\begin{equation}
\Lambda (\lambda,\lambda_1\cdots \lambda _M)=(\lambda +\eta)^N\prod_{\alpha =1}^{M}\frac{\lambda -v_{\alpha }-\eta}{\lambda -v_{\alpha}}+\lambda ^Nq^{M-1}\prod_{\alpha =1}^{M}\frac{\lambda -v_{\alpha }+\eta}{\lambda -v_{\alpha}}
\end{equation}
provided that
\begin{equation}
\left(\frac{v_{\alpha}+\eta}{v_{\alpha}}\right)^N=q^{M-1}\prod^{M}_{\beta \neq \alpha}\frac{v_{\alpha }-v_{\beta}+\eta}{v_{\alpha }-v_{\beta}-\eta}.
\end{equation}
Here $\alpha =1,\ldots,N$. 
If we perform a rescaling of the spectral parameter such that 
$v_{\alpha} \rightarrow v_{\alpha}/{\mathrm i} -{\eta}/{2}$ the energy spectrum is 
\begin{equation}
E=-\eta \sum^{M}_{\alpha =1}\frac{1}{v_{\alpha }+\eta ^2/4},
\end{equation}
where now the parameters $v_{\alpha }$ satisfy
\begin{equation}
\left(\frac{v_{\alpha}+{\mathrm i}\eta /2}{v_{\alpha}-{\mathrm i}\eta /2}\right)^N=q^{M-1}\prod^{M}_{\beta \neq \alpha}\frac{v_{\alpha }-v_{\beta}+{\mathrm i}\eta}{v_{\alpha }-v_{\beta}-{\mathrm i}\eta}.
\end{equation}
Note the way in which the anyonic  grading parameter $q$ appears in the Bethe ansatz
equations. It results in different distributions for the $v_{\alpha }$
than those for the standard $XXX$ model, leading to subtle physical properties \cite{LIM}.

\section{The  $t-J$ model of hard-core anyons}
\label{sec4}

Much work has been devoted during the last few decades towards a better
understanding of integrable models of strongly correlated electrons.
There are two kinds of  prototypical $t-J$ models \cite{azr} which are integrable -- 
the  integrable  {\it supersymmetric} $t-J$ model \cite{ffk,schlot} and the 
 integrable su(3) $t-J$ model \cite{Schl2,Zhou}.
Here we present an integrable su(3) $t-J$ model of interacting hard-core
anyons related to anyonic grading.  
The Hamiltonian reads
\begin{eqnarray}
\eta H &= &t\sum _{j=1}^{L}\sum_{\alpha=\uparrow,\downarrow}\left(\tilde{a}_{j\alpha}^{\dagger}\tilde{a}_{j+1,\alpha}+h.c.\right)\nonumber\\
& &
+J\left\{\sum _{j=1}^{L} \vec{S}_j\cdot \vec{S}_{j+1}+\frac{1}{4}n_jn_{j+1}\right\} +\sum _{j=1}^{L}(1-n_j)(1-n_{j+1}), \label{model2}
\end{eqnarray}
where
$n_j =n_{j\uparrow}+n_{j\downarrow}$ is the
number operator of single hard-core anyons with up and down spins.
Here $\tilde{a}_{j\alpha}^{\dagger}=a_{j\alpha}^{\dagger}(1-n_{j,-\alpha})$, which prohibits double occupancy. The model is integrable when  $J=2$ and $t=1$.
These operators satisfy the commutation relations
\begin{eqnarray}
& &\left\{a_{j\alpha},a_{j\alpha}\right\}=\left\{a_{j\alpha}^{\dagger},a_{j\alpha}^{\dagger}\right\}=0~~
\left\{a_{j\alpha},a^{\dagger}_{j\alpha}\right\}=1,\\
& &
a^{\dagger}_{i\downarrow}a_{j\downarrow}=q_1a_{j\downarrow}a^{\dagger}_{i\downarrow},~~a^{\dagger}_{i\uparrow}a_{j\uparrow}=q_2a_{j\uparrow}a^{\dagger}_{i\uparrow},\nonumber\\
& &a^{\dagger}_{i\uparrow}a_{j\downarrow}=q_3a_{j\downarrow}a^{\dagger}_{i\uparrow},~~a^{\dagger}_{i\downarrow}a_{j\uparrow}=q_3a_{j\uparrow}a^{\dagger}_{i\downarrow},\nonumber\\
& &a^{\dagger}_{j\downarrow}a^{\dagger}_{i\downarrow}=q_1a^{\dagger}_{i\downarrow}a^{\dagger}_{j\downarrow},~~a^{\dagger}_{j\uparrow}a^{\dagger}_{i\uparrow}=q_2a^{\dagger}_{i\uparrow}a^{\dagger}_{j\uparrow},\nonumber\\
&
&a^{\dagger}_{j\downarrow}a^{\dagger}_{i\uparrow}=q_3a^{\dagger}_{i\uparrow}a^{\dagger}_{j\downarrow},~~a^{\dagger}_{j\downarrow}a^{\dagger}_{i\uparrow}=q_3a^{\dagger}_{i\downarrow}a^{\dagger}_{j\uparrow},\nonumber\\
&
&q_1q_3^{-1}a^{\dagger}_{j\uparrow}a_{j\downarrow}a^{\dagger}_{i\downarrow}a_{i\uparrow}=q_3q_2^{-1}a^{\dagger}_{i\downarrow}a_{i\uparrow}a^{\dagger}_{j\uparrow}a_{j\downarrow},\nonumber\\
&& q_3q_1^{-1}a^{\dagger}_{i\uparrow}a_{i\downarrow}a^{\dagger}_{j\downarrow}a_{j\uparrow}=q_2q_3^{-1}a^{\dagger}_{j\downarrow}a_{j\uparrow}a^{\dagger}_{i\uparrow}a_{i\downarrow},
\end{eqnarray}
where $i>j$ is assumed. The spin operator is denoted by
$\vec{S}=\frac{1}{2}a_{\alpha}^{\dagger}{\vec{\sigma}}_{\alpha \beta}
c_{\beta}$, i.e. 
\begin{equation}
S^+=a^{\dagger}_{\uparrow}a_{\downarrow},\, S^-=a^{\dagger}_{\downarrow}a_{\uparrow},\,\,S^z=\frac{1}{2}(n_{\uparrow}-n_{\downarrow}).
\end{equation}
However, the spin exchange interaction is given by 
\begin{equation}
\vec{S}_j\cdot \vec{S}_{j+1}=\frac{1}{2}\left(\frac{q_3}{q_1}
S^{+}_{j+1}S^{-}_{j}+\frac{q_1}{q_3}
S^{+}_{j}S^{-}_{j+1}\right)+S_j^zS^z_{j+1}
\end{equation}
which evidently depends on the commutation parameters of the hard-core anyons.
Of course, we can present another  equivalent form of the spin exchange terms,
\begin{equation}
\vec{S}_j\cdot \vec{S}_{j+1}=\frac{1}{2}\left(\frac{q_3}{q_2}
S^{+}_{j+1}S^{-}_{j}+\frac{q_2}{q_3}
S^{+}_{j}S^{-}_{j+1}\right)+S_j^zS^z_{j+1},
\end{equation}
with the operators
\begin{equation}
S^+=a^{\dagger}_{\downarrow}a_{\uparrow},\, S^-=a^{\dagger}_{\uparrow}a_{\downarrow},\,\,S^z=\frac{1}{2}(n_{\downarrow}-n_{\uparrow}).
\end{equation}
In the above $q_i,~i=1,2,3$ are arbitrary anyonic parameters with the property
$q_i^{*}=q_i^{-1}$ for the Hamiltonian (\ref{model2}) to remain 
hermitian. In this model, on-site interaction between the hard-core
anyons preserves the Pauli exclusion principle. But anyonic phases
associated with the exchange of two particles at different sites
depend on their positions.  We also see that anisotropic spin exchange
interaction in the hard-core anyon su(3) $t-J$ model replaces the
ferromagnetic spin exchange in the standard su(3)  $t-J$
model \cite{Schl2,Zhou}. These free parameters act as anisotropic parameters
characterizing the anyon spin interaction. They lead to new phase
factors in the Bethe ansatz equations.

In order to link the anyonic  grading  $t-J$ model to Hamiltonian (\ref{model2}), 
we need to employ the $su(3)$ $R$-matrix \cite{schlot}
\begin{equation}
\stackrel{\vee}{R}(u)=\left(\begin{array}{ccccccccc}
a(u)&0&0&0&0&0&0&0&0\\
0&c(u)&0&b(u)&0&0&0&0&0\\
0&0&c(u)&0&0&0&b(u)&0&0\\
0&b(u)&0&c(u)&0&0&0&0&0\\
0&0&0&0&a(u)&0&0&0&0\\
0&0&0&0&0&c(u)&0&b(u)&0\\
0&0&b(u)&0&0&0&c(u)&0&0\\
0&0&0&0&0&b(u)&0&c(u)&0\\
0&0&0&0&0&0&0&0&a(u)
\end{array}\right)
\end{equation}
where
$$a(u)=u+\eta,\, \, \,b(u)=u,\, \,  \,  c(u)=\eta.$$
Given   the anyonic  grading 
\begin{eqnarray}
& &w(1,1)=q_1,\,\,\,w(2,2)=q_2,\,\,\,w(1,2)=w(2,1)=q_3,\nonumber\\
& &
w(3,1)=w(1,3)=w(3,2)=w(2,3)=w(3,3)=1\nonumber\\
\end{eqnarray}
one can show that the Lax operator
\begin{equation}
L_j(u)=\left(\begin{array}{ccc}
\begin{array}{l}q_1(\eta+u)n_{j\downarrow}\\
+u(q_3n_{j\uparrow}+1-n_j)\end{array}&q_3\eta a^{\dagger}_{j\uparrow}a_{j\downarrow}&\eta(1-n_{j\uparrow})a_{j\downarrow}\\
q_3\eta a^{\dagger}_{j\downarrow}a_{j\uparrow}&
\begin{array}{l}
u(q_3n_{j\downarrow}+1-n_j)\\
+q_2(u+\eta)n_{j\uparrow}\end{array}&
\eta(1-n_{j\downarrow})a_{j\uparrow}\\
\eta a_{j\downarrow}^{\dagger}(1-n_{j\uparrow})&
\eta a_{j\uparrow}^{\dagger}(1-n_{j\downarrow})&
u+\eta(1-n_j)\end{array}\right)
\end{equation}
generates the local anyonic grading YBE (\ref{CYBE4}). 
As a consequence,  the integrals of motion of
the model can be obtained from the expansion of the
transfer matrix in the spectral parameter $u$. Using expressions
(\ref{Ham}) and (\ref{Hamd}) with the above anyonic  grading, 
the Hamiltonian (\ref{model2}) can be derived from the relation 
\begin{equation}
\tau (\lambda)=(1+H\lambda +\cdots )\tau(0).
\end{equation}
In this way the integrability of the supersymmetric $t-J$ model of hard-core
anyons (\ref{model2}) is guaranteed by the anyonic grading  Yang-Baxter equations (\ref{CYBE}). 
Special choices of the grading parameters characterize different statistical mechanical models. 
For example, if   $q_i=1$ the model becomes the $su(3)$ Heisenberg model \cite{schlot} in
terms of hard-core bosons \cite{JW2}.  Whereas if $q_i=-1$ the model 
becomes a $su(3)$ $t-J$ model \cite{Schl2,Zhou}. We see then that these parameters characterize
different statistics and will result in different physical properties. 
We now turn to the nested algebraic Bethe ansatz
\cite{ffk} to derive the exact solution of the model.

Define
\begin{equation}
T(u) = L_L(u)\cdots L_1(u)=
\left(\matrix{A_{11}(u)&A_{12}(u)&B_1(u)\cr
A_{21}(u)&A_{22}(u)&B_{2}(u)\cr
C_1(u)&C_{2}(u)&D(u)\cr }\right)
\end{equation}
acting on the anyonic  Hilbert space.
We choose the vacuum state $|0\rangle=\prod_{i=1}^{L}\otimes_{{\rm
    c}}|0\rangle_i$ with $a_{i\alpha}|0\rangle =0$. 
The nested algebraic Bethe ansatz solution of the usual
$t-J$ model has been discussed at length in the literature \cite{ffk}, 
so here we highlight only the differences in the nesting structure
caused by the anyonic grading.
The commutation relations between the diagonal
fields and the creation fields are
\begin{eqnarray}
D(u_1)C_a(u_2) &= &
\frac{a(u_2-u_1)}{b(u_2-u_1)}C_a(u_2)D(u_1)
-\frac{c(u_2-u_1)}{b(u_2-u_1)}C_a(u_1)D(u_2)\label{com1}\\
A_{ab}(u_1)C_f(u_2)& =&  \frac{a(u_1-u_2)}{b(u_1-u_2)}\left\{r^{(1)}(u_1-u_2)^{bf}_{ed}w(e,a)C_e(u_2)A_{ad}(u_1)\right\} \nonumber\\
& &
-\frac{c(u_1-u_2)}{b(u_1-u_2)}w(b,a)C_b(u_1)A_{ac}(u_2)\label{com2}
\end{eqnarray} 
with
\begin{eqnarray}
& &r^{aa}_{aa}=1,\,a=1,2,\,\,\, r^{ab}_{ab}=\frac{c(u)}{a(u)},\,a\neq b=1,2,\nonumber\\
& &r^{ab}_{ba}=\frac{b(u)}{a(u)},\,a\neq b=1,2. \label{r1}
\end{eqnarray}
The anyonic  grading functions appearing in the commutation relation (\ref{com2}) are kept in the 
nested transfer matrix for the spin degree of freedom. 
This makes the nested algebraic Bethe ansatz very complicated.  
We see however, that the first term in each of the commutation relations
(\ref{com1})--(\ref{com2}) contribute to the eigenvalues of the
transfer matrix which should be analytic functions of the spectral
parameter $u$.  Consequently, the residues at singular points must
vanish. This yields the Bethe-ansatz equations which in turn assure
the cancellation of the unwanted terms in the eigenvalues of the
transfer matrix. To this end, we  choose Bethe state
$|\Phi\rangle$ as
\begin{eqnarray}
|\Phi \rangle =C_{g_1}(u_1)\cdots C_{g_{N}}(u_{N})|0\rangle F^{g_N\cdots g_1}. \label{n-state}
\end{eqnarray}
Following the standard procedure of the algebraic Bethe ansatz, 
the eigenvalue of the monodromy matrix acting on the state
(\ref{n-state}) is obtained as 
\begin{eqnarray}
& &\tau(u)|\Phi\rangle = \Lambda (u,\{u_i\})|\Phi\rangle = (u+\eta)^L\prod_{i=1}^{N}\frac{(u-u_i-\eta)}{(u-u_i)}|\Phi\rangle\nonumber\\
& &
+u^L\prod^N_{i=1}\frac{u-u_i+\eta}{u-u_i}\prod_{l=1}^{N}C_{gl}(u_l)|0\rangle\left[\tau^{(1)}(u)\right]^{h_1\cdots h_N}_{g_1\cdots g_N}F^{g_N\cdots g_1}
\end{eqnarray}
provided that 
\begin{eqnarray}
\frac{(u+\eta)^L}{u^L}\prod^{N}_{\stackrel{\scriptstyle l=1}{l\neq i}}\frac{u_i-u_l-\eta}{u_i-u_l+\eta}=\left[\tau^{(1)}(u)\right]^{h_1\cdots h_N}_{g_1\cdots g_N}\mid _{u=u_i}.
\end{eqnarray}
In the above the nested transfer matrix is given by 
\begin{eqnarray}
\left[\tau^{(1)}(u)\right]^{h_1\cdots h_N}_{g_1\cdots g_N}&=&
{\rm atr}_0\left(L^{(1)}_N(u-u_N)^{d_{N-1}g_N}_{ah_N}L^{(1)}_{N-1}(u-u_{N-1})^{d_{N-2}g_{N-1}}_{d_{N-1}h_{N-1}}\right.\nonumber\\
& &
\left.
\cdots L^{(1)}_2(u-u_2)^{d_{1}g_2}_{d_2h_2}L^{(1)}_1(u-u_1)^{ag_1}_{d_1h_1}\right).
\end{eqnarray}
where the local Lax operator reads
\begin{equation}
L^{(1)}_j(u)=\left(\begin{array}{cc}
q_1n_{\downarrow}+\frac{b(u)}{a(u)}q_3n_{\uparrow}&\frac{c(u)}{a(u)}q_3a^{\dagger}_{\uparrow}a_{\downarrow}\\
\frac{c(u)}{a(u)}q_3a^{\dagger}_{\downarrow}a_{\uparrow}&q_2n_{\uparrow}+\frac{b(u)}{a(u)}q_3n_{\downarrow}\end{array}\right).\label{LR2}
\end{equation}
This satisfies the anyonic grading YBE (\ref{gYBA}) with grading $ w(1,1)=q_1,\,w(2,2)=q_2,\,w(1,2)=w(2,1)=q_3$. 
This realization of the nested Lax operator (\ref{LR2}) paves the 
way to diagonalize the transfer matrix of the model. 
After some algebra, we obtain the eigenvalue of the transfer matrix in the form
\begin{eqnarray}
& & \Lambda (u,\{u_i\}\{v_i\})= (u+\eta)^L\prod_{i=1}^{N}\frac{(u-u_i-\eta)}{(u-u_i)}+u^Lq_1^{M-1}q_3^{N-M}\prod^{M}_{l=1}\frac{u-v_l+\eta}{u-v_l}\nonumber\\
& &
+u^Lq_2^{N-M-1}q_3^M\prod^{M}_{i=1}\frac{u-u_i+\eta}{u-u_i}\prod^{M}_{l=1}\frac{u-v_l-\eta}{u-v_l}.
\end{eqnarray}
Here the quantum numbers $N$ and $M$ are the total number of hard-core anyons and the number of 
hard-core anyons with down spin, respectively.
The parameters $u_i$ and $v_l$ characterize the charge and spin rapidities of the model.
If making a rescaling $u_i\rightarrow u_i-\eta/2$, $ v_i\rightarrow v_i-\eta$,
the Bethe ansatz equations are given by 
\begin{eqnarray}
& &\frac{(u_i+\frac{{\mathrm i}\eta}{2})^L}{(u_i-\frac{{\mathrm i}\eta}{2})^L}=q_2^{N-M-1}q_3^{M}\prod_{\stackrel{\scriptstyle l=1}{l\neq i}}^{N}\frac{u_i-u_l+{\mathrm i}\eta}{u_i-u_l-{\mathrm i}\eta}\prod^{M}_{l=1}\frac{u_i-v_l-\frac{{\mathrm i}\eta}{2}}{u_i-v_l+\frac{{\mathrm i}\eta}{2}},\\
& &
q_1^{M-1}q_2^{-(N-M-1)}q_3^{N-2M}\prod ^{N}_{i=1}\frac{v_j-u_i-\frac{{\mathrm i}\eta}{2}}
{v_j-u_i+\frac{{\mathrm i}\eta}{2}}=\prod^{M}_{\stackrel{\scriptstyle l=1}{l\neq j}}\frac{v_j-v_l-{\mathrm i}\eta}{v_j-v_l+{\mathrm i}\eta},\nonumber
\end{eqnarray}
for $i=1,\ldots ,N$ and $j=1,\ldots ,M$.
In this way we have the energy spectrum 
\begin{equation}
E=L-\eta^2\sum^{N}_{i=1}\frac{1}{u_i^2+\frac{\eta^2}{4}}.
\end{equation}

\section{Conclusion}
\label{sec5}

In summary, we have constructed a class of integrable models
associated with anyonic grading. We found that these integrable models
may be used to describe the interaction of hard-core anyons. With
regard to the anyonic grading supersymmetric structure, we presented a
unifying approach -- the anyonic grading QISM -- to treat this class
of integrable models. We explicitly constructed integrable models of
hard-core anyons associated with the $XXX$ model and the $t-J$ model
with anyonic grading. The exact solutions of these models were
obtained by means of the algebraic Bethe ansatz. It is seen that the
phase functions associated with the exchange of two hard-core anyons
at different sites lead to nontrivial phase factors in the Bethe
ansatz equations. These phase factors encode the anyonic effects in
these models.  Application of the anyonic grading QISM to the 1D
Hubbard model \cite{Hubbardbook} and the su(2,1) supersymmetric $t-J$ model
with different gradings would provide an interesting generalization of
strongly correlated systems \cite{Zhou,gradings} to models of hard-core anyons. 
We hope to consider these problems and their
ground state properties elsewhere.

\ack 
The authors thank the Australian Research Council for support.
A Foerster thanks CNPq (Conselho Nacional de Desenvolvimento Cient\'{\i}fico e Tecnol\'ogico) and 
FAPERGS (Funda\c{c}\~{a}o de Amparo \~{a} Pesquisa do Estado do Rio
Grande do Sul) for financial support. 
H-Q Zhou acknowledges support from the National
Natural Science Foundation of China (grant No. 10774197)
and from the Natural Science Foundation of Chongqing
(CSTC2008BC2023).

\end{document}